\documentstyle{article}
\include{epsf}
\begin{document}
\title{A Geometrical Description of the Consistent and Covariant Chiral Anomaly}
\author{{\bf C.Ekstrand}\\ Department of Theoretical Physics, \\ Royal Institute of
Technology, \\ S-100 44 Stockholm, Sweden \\ ce@theophys.kth.se}
\date{}
\maketitle
\newcommand{\eq}{\begin{equation}}
\newcommand{\eqend}{\end{equation}}
\newcommand{\eqa}{\begin{eqnarray}}
\newcommand{\eqaend}{\end{eqnarray}}
\newcommand{\nonu}{\nonumber \\ \nopagebreak}
\newcommand{\pow}{^}
\newcommand{\Ref}[1]{(\ref{#1})}
\newcommand{\B}{{\cal B}}
\newcommand{\bbar}{\bar}
\newcommand{\F}{{\cal F}}
\newcommand{\cL}{{\cal L}}
\newcommand{\ep}{{\cal E}}
\newcommand{\EE}{{\mbox{\scriptsize\bf E}}}
\newcommand{\E}{{\bf E}}\newcommand{\SC}{{\bf{A}}}
\newcommand{\CU}{{\bf{F}}}
\newcommand{\W}{{\cal W}}
\newcommand{\M}{\cal M}
\newcommand{\Ss}{S}
\newcommand{\A}{{\cal A}}
\newcommand{\T}{\cal T}
\newcommand{\D}{\cal D}
\newcommand{\g}{\cal g}
\newcommand{\G}{{\cal G}}
\newcommand{\Rr}{\cal R}
\newcommand{\Pp}{\cal P}
\newcommand{\etta}{X}
\newcommand{\C}{\cal C}
\newcommand{\V}{\cal V}
\newcommand{\mmu}{\mu}
\newcommand{\mmmu}{\lambda ^{\prime}}
\newcommand{\mdim}{{2n}}
\newcommand{\ddd}{d}
\newcommand{\uu}{u}
\newcommand{\ommega}{\omega ^{\prime}}
\newcommand{\ommmega}{\omega }
\newcommand{\nejnum}{\nonumber}
\def\eop{\nopagebreak\hfill $\Box$}
\newtheorem{definition}{Definition}
\newtheorem{lemma}{Lemma}
\newtheorem{theorem}{Theorem}
\newtheorem{proposition}{Proposition}
\newtheorem{corollary}{Corollary}
\begin{abstract}
A geometrical interpretation of the consistent and covariant chiral anomaly is done in the space-time 
respective Hamiltonian framework. 
\end{abstract}
\section{Introduction}
A mathematical rigorous definition of the consistent anomaly in the space-time formalism is in terms of a connection on a line bundle over $\A$, the space of gauge 
potentials. Similarly, the consistent Schwinger term is defined by a curvature on 
a line bundle over a dense subset of $\A$. Despite this, no geometrical description has been given for the covariant counterparts. In this paper we will obtain a geometrical interpretation of the covariant 
anomaly and Schwinger term in a similar way to what has been done in the 
consistent framework. For completeness, we will also review the consistent case. The result is that the consistent and covariant anomaly are given by 
connections on line bundles. The consistent Schwinger term is given by the 
curvature of a line bundle while for the covariant Schwinger term this is 
only true up to a (canonical) form defined by data coming from a part of  
space-time which are far from the point of interest. Explicit expressions 
for the anomalies and Schwinger terms have been given in \cite{AES}, where the correct expression for the covariant Schwinger term was determined (there are two results in the literature, differing by a sign). The consistent and covariant descent equations will also be compared. The connecting term between consistent and covariant cochains is shown to be a local form on the space of gauge potentials.

The paper is organized in the following way: In section 2 and 3 we consider 
the anomaly in the space-time respective Hamiltonian framework. In section 4 
we review the descent equations and compare the consistent and covariant cochains. 

\section{The chiral anomaly in Euclidean space-time}
We will consider Weyl fermions coupled to an external gauge field $A\in\A$ in a $2n$-dimensional space-time $M$, a smooth, flat, compact and oriented Riemannian spin manifold without boundary. The group $\G$ of gauge transformations consists of diffeomorphisms $\varphi $ of a principal bundle $\pi :P\stackrel{G}{\rightarrow }M$  such that the base remains unchanged. It acts on the affine space $\A$ of connections on $P$ by pull-back: $A\cdot \varphi :=\varphi^\ast A$. To make the action free (so $\A /\G$ will be a smooth manifold) we will assume that $\G$ only consists of diffeomorphisms that leaves a fixed point $p_0\in P$ unchanged. 

The generating functional is defined as
\[
\exp (-W(A))=\int d\psi d\bar{\psi }\exp (-\int _M\bar{\psi }\partial \!\!\! /_A^+\psi d^{2n}x),
\]
where $W$ is the effective action and $\partial \!\!\! /_A^+ =\partial \!\!\! /_A(1+\gamma _5)/2=\gamma ^\mu (\partial _\mu +A_\mu )(1+\gamma _5)/2$. We will use conventions such that $\gamma ^\mu $ is hermitian and $A_\mu$ is anti-hermitian. The consistent chiral anomaly measures the lack of gauge invariance of the generating functional:
\eq
\label{eq:0}
\delta _X\exp (-W(A))=-\exp (-W(A))\omega (A;X),
\eqend
$X\in \mbox{Lie}\G$. The reason for the minus sign is that the consistent anomaly is defined by $\delta _XW(A)=\omega (A;X)$. The effective action is however only defined up to terms $c$ which are local functionals in $\A$. This makes the generating functional defined up to $\exp (-c(A))$ and the anomaly up to terms $(\delta _Xc)(A)$.

 To make mathematical sense out of the effective action we will first assume that $\partial \!\!\! /_A \pow +$ is a Fredholm operator with zero index. We can then construct the canonical section $\mbox{det}i\partial \!\!\! /_A$ of the determinant line bundle $\mbox{DET}i\partial \!\!\! /_A =\mbox{det ker}i\partial \!\!\! /_A^+\otimes (\mbox{det coker}i\partial \!\!\! /_A^+)^\ast$, see \cite{BGV}. To identify it with a functional, a reference section $s_0$ is chosen. Then we define $\exp (-W(A))$ as $\mbox{det}i\partial \!\!\! /_A/s_0$. Since all line bundles over an affine space are equivalent, it might appear meaningless to consider the quotient of two sections of a line bundle over $\A$. However, the determinant line bundle is not just a line bundle, but it is also equipped with a certain structure. For instance, below we will review the fact that the restriction to gauge directions (vectors tangent to the fibres in $\A\rightarrow \A /\G$) of its connection (covariant derivative) $\nabla $ can be determined canonically.

 The variation $\delta _X$ will be defined with respect to a fixed $s_0$:
\[
\delta _X\exp (-W(A)):=(\nabla _X\mbox{det}i\partial \!\!\! /_A)/s_0.
\]
Notice that it in general is not possible to find a section such that $\nabla _Xs_0(A)=0$ for all $A\in\A$. Thus, if we consider a variation at a point $A\pow \prime $ which is not related to $A$ by a gauge transformation, then a different section $s_0\pow \prime$ has to be chosen. 
Relation \Ref{eq:0} is now equivalent with 
\eq
\label{eq:1}
\nabla _X\mbox{det}i\partial \!\!\! /_A=-\mbox{det}i\partial \!\!\! /_A\omega (A;X) .
\eqend
Observe that this definition of the anomaly does not depend on the choice of $s_0$.

 Recall that a connection on a line bundle satisfies the Leibniz rule $\nabla (s\lambda )=(\nabla s)\lambda +sd\lambda $ for any section $s$ and function $\lambda $. The connection can be described by a 1-form $\omega $ on $\A$ by $\nabla s=s\omega $, where $s$ is a nowhere vanishing section. We will refer to $\omega $ as the pull-back of $\nabla $ with respect to $s$. Observe that $\omega $ depends on the choice of pull-back: if $\omega ^\prime$ is defined with respect to $s^\prime =s\lambda$, then $ \omega ^\prime =\omega +\lambda ^{-1}d\lambda $. For $\lambda = e^c$ it gives: $\omega ^\prime(A;X)=\omega (A;X)+(\delta _Xc)(A)$, where $X$ in this case is any vector field on $\A$. From this abstract discussion it follows that if locality is disregarded, then the anomaly can be identified with minus the restriction to gauge directions of the connection on the determinant line bundle. When locality is taken into account, we obtain the following identification: The anomaly is minus the restriction to gauge directions of the pull-back of the connection with respect to sections given by $\mbox{det}i\partial \!\!\! /_A$ multiplied with the exponential of local terms.

 The curvature (pulled-back to $\A$) corresponding to $\nabla$ has been computed in \cite{BF} (we will suppress the representation of the gauge group): 
\[
F=c_n\int _M\mbox{tr}\left( \F ^{n+1}\right) ,\quad c_n=-2\pi i \frac{1}{(n+1)!}\left(\frac{i}{2\pi }\right) ^{n+1}. 
\]
Here, $\F =(d+\delta)\alpha +\alpha ^2$ is the curvature corresponding to a connection on the universal bundle $P\times\A\rightarrow M\times \A$. We thus see that the curvature (and connection) of the determinant line bundle depends on the choice of $\alpha $. We will soon see how it is possible to partly determine $\alpha $ by physical arguments. This will lead us to an expression for the anomaly. Notice that in order to compute the anomaly from $F$ we need to use the fact that the anomaly (the consistent and the covariant) is local. This since there certainly exist forms with $\delta \omega =F$ which are non-local even in gauge directions. 

The Dirac operator acts on sections of an associated bundle to $P\rightarrow M$. In the family index theorem one then consider the universal bundle as bundles $P\rightarrow M$ parameterized by $\A$. The first reasonable choice of $\alpha $ is then $\alpha =A$. In general, the bundle $P\rightarrow \A$ may twist over $\A$. This means that $\alpha $ is of the form $A+a$, where $a$ is a 1-form on $\A$ taking values in $\mbox{Lie}G$. Such a twist occurs for example when demanding gauge invariance, i.e. that $A+a$ should descend to a connection on $(P\times \A )/\G \rightarrow M\times \A /\G$, where $\G$ acts on $P\times\A$ by $(p,A)\cdot \varphi :=(\varphi ^{-1}(p),A\cdot \varphi )$, \cite{AS}. This is equivalent with that the determinant line bundle can be pushed forward to a line bundle on $\A /\G$ (where it is equipped with a curvature). This does not determine $a$ uniquely, but it implies that it is equal to the Faddeev-Popov ghost $v$ in gauge directions (one example is the non-local form $a=(d_A^\ast d_A)^{-1}d_A^\ast $). This gives the consistent anomaly. 

\pagebreak

To compute the anomaly, recall that 
\eqa
\label{eq:2}
\mbox{tr}\left(\F ^{n+1}\right)-\mbox{tr}\left(\F ^{\prime n+1}\right) & = & (d+\delta )\omega _{2n+1}(\alpha  ,\alpha ^\prime )\nonu
\omega _{2n+1}(\alpha  ,\alpha ^\prime ) & = & (n+1)\int _0^1dt\mbox{tr}\left( (\alpha  -\alpha ^\prime )\F _t^n\right)
\eqaend
and $\F _t$ the curvature corresponding to $(1-t)\alpha ^\prime  +t\alpha $. We will first assume that $P$ is trivial so that $\alpha ^\prime =0$ is a possible choice. With $\omega $ the pull-back of $\nabla $, $F=\delta \omega $, we get 
\[
\omega =c_n\int _M\omega _{2n+1}(A+a ,0)+\delta\chi ,
\]
where $\chi$ is an arbitrary functional. Let us now restrict to gauge directions. Recall that the exterior differential $\delta $ then is equal to the  BRST operator. The first term on the right hand side is now local. By restricting to local $\chi$'s, the right hand side becomes minus the consistent anomaly. The Russian formula is the fact that $\F =dA+A^2$ (in gauge directions). It implies that $F =\delta \omega $ is zero in gauge directions. This is the Wess-Zumino consistency condition. It implies that it is possible to regard the consistent anomaly as an element in a cohomology group.

We will now get rid of two unwanted assumptions, namely that $\mbox{ind}\partial \!\!\! /_A \pow +=0$ and that $P$ is trivial. For this reason we consider the \lq difference\rq $ $ between two Dirac operators: $i\partial \!\!\! /_A \pow ++(i\partial \!\!\! /_{A\pow \prime } \pow +)\pow \ast =i(\partial \!\!\! /_A \pow ++\partial \!\!\! /_{A\pow \prime } \pow -)$. This is a Dirac operator with zero index (since $A \pow \prime$ is a connection on the same bundle $P$) and we can thus study the section $\mbox{det}i(\partial \!\!\! /_A \pow ++\partial \!\!\! /_{A\pow \prime } \pow -)$ of  $\mbox{DET}i(\partial \!\!\! /_A \pow ++\partial \!\!\! /_{A\pow \prime } \pow -)\rightarrow \A\times\A$, where $A$ is in the first factor of $\A\times\A$ and $A\pow \prime $ is in the second. 
The curvature on $\mbox{DET}i(\partial \!\!\! /_A \pow ++\partial \!\!\! /_{A\pow \prime } \pow -)$ is 
\eq
\label{eq:3}
F-F^\prime =c_n\int _M\mbox{tr}\left( \F ^{n+1} -\F ^{\prime n+1}\right) =(\delta +\delta ^\prime )c_n\int _M\omega _{2n+1}(A+a,A^\prime +a^\prime ) .
\eqend
Since the consistent anomaly is defined with respect to variations of $A$, it is 
obtained by only considering form parts with respect to the first factor of $\A\times\A$. With $F=\delta \omega $ we then get:
\[
\omega =c_n\int _M\omega _{2n+1}(A+a ,A^\prime )+\delta\chi ,
\]
where $\chi$ is local when restricted to gauge directions. The second determinant bundle has thus only been used as a reference. The gauge field $A^\prime$ is unaffected by the variations we are interested in. The anomaly above is referred to as the consistent anomaly in the background connection $A^\prime$, \cite{MSZ}. Expressions obtained from two different background connections differ by a BRST operator acting on a local term and represents therefore the same anomaly. Notice that since we let everything be fixed in the second factor, we do not need to demand gauge invariance of $\alpha ^\prime$. We may thus set $a^\prime =0$. The corresponding canonical section and line bundle are then no longer symmetric in $A$ and $A\pow \prime $ and we will for this reason denote them by $\widehat{\mbox{det}}i(\partial \!\!\! /_A \pow ++\partial \!\!\! /_{A\pow \prime } \pow -)$ respective $\widehat{\mbox{DET}}i(\partial \!\!\! /_A \pow ++\partial \!\!\! /_{A\pow \prime } \pow -)$. Notice that since $A\pow \prime $ is regarded as fixed,  $\widehat{\mbox{DET}}i(\partial \!\!\! /_A \pow ++\partial \!\!\! /_{A\pow \prime } \pow -)$ can be considered as a line bundle over $\A$ (the first factor of $\A\times\A$) with section $\widehat{\mbox{det}}i(\partial \!\!\! /_A \pow ++\partial \!\!\! /_{A\pow \prime } \pow -)$. 

Let us make a comment on the line bundle $\mbox{DET}i\partial \!\!\! /_{A}\pow +\otimes(\mbox{DET}i\partial \!\!\! /_{A^\prime }\pow + )^\ast $. Despite the fact that it has the same curvature as $\mbox{DET}i(\partial \!\!\! /_A \pow ++\partial \!\!\! /_{A\pow \prime } \pow -)$, it is equipped with a different structure. For instance, it is easy to see that its connection evaluated on the canonical section $\mbox{det}i\partial \!\!\! /_{A}\pow +\otimes(\mbox{det}i\partial \!\!\! /_{A^\prime }\pow +)^\ast $ gives a different result than for $\mbox{det}i(\partial \!\!\! /_A \pow ++\partial \!\!\! /_{A\pow \prime } \pow -)$. In the first case one obtains a form which consist of two separate pieces on the first respective second factor of $\A\times\A$. However, in the second case we saw in the expression for the consistent anomaly that the background connection also plays a role in the first factor. This is an example of the multiplicative anomaly \cite{KO} (the determinant of a product is not necessary equal to the product of the determinants for the factors) in the family case, i.e. 
\begin{eqnarray*}
\mbox{DET}i\partial \!\!\! /_{A}\pow +\otimes\mbox{DET}(i\partial \!\!\! /_{A^\prime }^+) \pow \ast  & = & \mbox{DET}\left(\begin{array}{cc} 0 & 1\\ i\partial \!\!\! /_{A} \pow + & 0\end{array}\right)\otimes\mbox{DET}\left(\begin{array}{cc}0 & i\partial \!\!\! /_{A^\prime }^- \\ 1& 0 \end{array}\right)\nonu 
& \neq & \mbox{DET}\left(\begin{array}{cc} 0 & i\partial \!\!\! /_{A^\prime }^- \\ i\partial \!\!\! /_{A} \pow + & 0\end{array}\right) = \mbox{DET}i(\partial \!\!\! /_A \pow ++\partial \!\!\! /_{A\pow \prime } \pow -),
\end{eqnarray*}
where we with the (non-)equality mean with respect to line bundles with structures.

Consider now $\widehat{\mbox{DET}}i(\partial \!\!\! /_A \pow ++\partial \!\!\! /_{A\pow \prime } \pow -)$ and $\widehat{\mbox{det}}i(\partial \!\!\! /_A \pow ++\partial \!\!\! /_{A\pow \prime } \pow -)$
over the diagonal in $\A\times\A$, i.e. we put $A^\prime =A$. The curvature then looks as in eq. \Ref{eq:3} with $a ^\prime =0$ and $A^\prime =A$. The operator $\delta ^\prime$ acts as $\delta$ on $A^\prime =A$ in the second argument of $\omega _{2n+1}$ while it leaves $A$ and $a$ in the first argument unchanged. We then see that if we identify the diagonal with $\A$, then the bundle has the curvature 
\eq
\label{eq:4}
F=\delta c_n\int _M\omega _{2n+1}(A+a ,A),
\eqend 
where $\delta $ now acts on the second argument as well. The corresponding connection 
\[
\omega =c_n\int _M\omega _{2n+1}(A+a ,A)+\delta\chi 
\]
is recognized as (minus) the covariant anomaly if restricted to gauge directions (where $\chi $ is local).


\begin{picture}(120,170)(-80,-40)
\put(0,0){\line(0,1){100}}
\put(0,0){\line(1,0){100}}
\put(0,0){\line(1,1){100}}
\put(0,50){\line(1,0){100}}
\put(-10,50){$\A$}
\put(50,-10){$\A$}
\put(20,80){$\begin{array}{c} \widehat{\mbox{DET}}\\ \downarrow \\ \A\times\A\end{array}$}
\put(110,50){consistent}
\put(110,100){covariant}
\put(-50,-30){\footnotesize{}fig. 1: The choice made for the consistent respective covariant anomaly.}
\end{picture}

It is now clear that also in the covariant formalism can the anomaly be defined by eq. \Ref{eq:1} and the discussion that followed this equation, however, the connection and canonical section are defined with respect to $\widehat{\mbox{DET}}i(\partial \!\!\! /_A \pow ++\partial \!\!\! /_{A} \pow -)$. The fact that the curvature is non-zero in gauge directions means that the Wess-Zumino consistency condition is not fulfilled. It also implies that the covariant anomaly can not be obtained from a variation of a functional. For instance, since the curvature is non-zero also in gauge directions it is impossible to find a reference section satisfying $\nabla _Xs_0=0$ at a point $A\in\A$.


\section{The chiral anomaly in the Hamiltonian formalism}
We will here let $M$ be $(2n-1)$-dimensional and interpreted as the physical space at a fixed time. $\G$ and $\A$ will be the corresponding group of gauge transformations respective space of potentials. The chiral anomaly manifest itself as the Schwinger term in the Hamiltonian formalism. We will take the starting point that the consistent Schwinger term is minus the curvature of the vacuum bundle restricted to gauge directions. To compute it we must therefore understand the vacuum first. 

The Hamiltonian $H_A$ decomposes the 1-particle Hilbert space into two pieces, depending on if the eigenvalues are bigger or less than some number $\lambda$:
\[
H=H_+(A,\lambda )\oplus H_-(A,\lambda ).
\]
The vacuum bundle $\Omega _\lambda (A)$ is given by the filled up Dirac sea to a certain level $\lambda$. It is thus the complex span of the wedge product of the eigenvectors in $H_-(A,\lambda )$. Naively, this is a line bundle over $U_\lambda =\{ A\in \A  | \lambda \in \!\!\!\!\! |\, \mbox{spec}(H_{A})\}$. However, the vacuum bundle is not well-defined since it involves an infinite wedge product. Instead, we will consider the \lq quotient\rq $ $ of two vacua. For this, consider a decomposition of the 1-particle space with respect to different $A$ and $\lambda $: $H=H_+(A^\prime ,\lambda^\prime  )\oplus H_-(A^\prime ,\lambda^\prime  )$. The intuitive definition of the quotient of $\Omega _\lambda (A)$ and $\Omega _\lambda (A^\prime)$ is then $\mbox{det} H_-(A,\lambda )\otimes (\mbox{det} H_-(A^\prime ,\lambda^\prime  ))^\ast$, where $\mbox{det}$ is the top wedge product. This ill-defined expression becomes simpler if we quote out a common infinite part from the first and the second factor. The quotient of the two vacuum bundles can then be defined as $(\mbox{det}H_-(A,\lambda )\cap H_+(A^\prime ,\lambda^\prime  ))\otimes (\mbox{det}H_+(A,\lambda )\cap H_-(A^\prime ,\lambda^\prime  ))^\ast $. However, this is still not well-defined since the intersections are infinite dimensional spaces. Fortunately there exists renormalization methods to handle this problem. In 1 space dimension it is particular simple since the identity operator on $H$ becomes a Hilbert-Schmidt operator when restricted to the domain $H_-(A,\lambda )$ ($H_+(A,\lambda )$) and the range $H_+(A^\prime ,\lambda^\prime  )$ ($H_-(A^\prime ,\lambda^\prime  )$).

 Although the above definition of the quotient of two vacua is the most common used, we will use an alternative definition from \cite{CMM}. Instead of using the identity operator on $H$ to compare $H_-(A,\lambda )$ and $H_-(A^\prime ,\lambda^\prime  )$, the Dirac operator will be used. For this, let $A(t)$, $t\in I=[0,1]$ be a path in $\A$ from $A^\prime $ to $A$. It is often useful to consider $t$ as the time. By letting the wave functions $\psi$ depend on the extra parameter $t$, we see that the Dirac equation $i\partial \!\!\! /_{A(t)}^+\psi =0$, $i\partial \!\!\! /_{A(t)}^+=i\partial _t -H_{A(t)}$, can be used to identify the two differently composed Hilbert spaces. Then $\mbox{DET}_{\lambda\lambda^\prime }(A(t)):= \mbox{det ker}i\partial \!\!\! /_{A(t)}^+\otimes (\mbox{det coker}i\partial \!\!\! /_{A(t)}^+)^\ast$ is a reasonable definition of $\Omega _\lambda (A)\otimes\Omega _\lambda (A^\prime )^\ast$ if the boundary conditions $\psi |_{t=0}\in H_+(A^\prime ,\lambda^\prime  )$ and $\psi |_{t=1}\in H_-(A,\lambda )$ are used. This gives a line bundle over $U_\lambda \times U_{\lambda ^\prime }\subset \A\times\A$ in a similar way as for the 
anomaly in the space-time formalism. Its curvature is
\eq
\label{eq:5}
F=c_n\left(\int _{M\times I}\mbox{tr}\left( \F ^{n+1}\right) -\frac{1}{2}\left(\hat{\eta }_\lambda \right) _{[2]} +\frac{1}{2}\left(\hat{\eta }_{\lambda ^\prime } \right) _{[2]} \right),
\eqend
where $\F $ is the curvature corresponding to a connection $\alpha $ on the universal bundle $P\times I\times (U_\lambda \times U_{\lambda ^\prime })\rightarrow M\times I\times (U_\lambda \times U_{\lambda ^\prime })$. The fact that $M\times I$ is a manifold with boundary gives rise to additional contributions, the 2-form parts of the $\hat{\eta }$-forms, \cite{APS,PZ2}.

Let us now compute the consistent Schwinger term from the above results. We thus want to compute $F$ when gauge variations of $A=A(t=1)$ is made. For this reason, we chose the simplest possible path, namely 
$A(t)=(1-t)A\pow {\prime }+tA$ (actually, in order to use theorems about 
determinant line bundles, $t$ should be replaced with a function $f(t)$ 
which is 0 in a neighbourhood of $t=0$ and 1 in a neighbourhood of $t=1$, 
\cite{EM}). Let us now regard $A\pow {\prime }$ as 
a background connection. Then $\mbox{DET}_{\lambda \lambda \pow {\prime 
}}(A(t))$ is a line bundle over $U_\lambda \subset \A$. Clearly, the choice $\alpha 
=(1-t)A\pow {\prime }+t(A+a)$ should be made, where $a$ is equal to the 
ghost in gauge directions. It implies that $\F =(d+d_t+\delta )((1-t)A\pow 
{\prime }+t(A+a))+((1-t)A\pow {\prime }+t(A+a))\pow 2$. By dimensional 
reasons, only one $d_t$ term can appear in $\int _{M\times I}\mbox{tr}\left( 
\F \pow {n+1}\right) $. It gives
\eq
\label{eq:6}
\int _{M\times I}\mbox{tr}\left( \F \pow {n+1}\right) = \int _M\omega _{2n+1 
}(A+a,A\pow {\prime }).
\eqend
To compute the consistent Schwinger term, we now restrict eq. \Ref{eq:5} to 
gauge directions. By construction, the $\hat{\eta }$-forms vanishes then, 
\cite{CMM}. Thus, the consistent Schwinger term in a background connection 
$A\pow {\prime }$ is given by minus the restriction of eq. \Ref{eq:6} to 
gauge directions.

We have shown that the vacuum bundle can be defined as a determinant bundle. 
As in the space-time formalism it is not the bundles themselves that are 
interesting on the level of anomalies, but the fact that the determinant line bundle is equipped with 
a natural connection in gauge directions. In fact, regarding Schwinger 
terms, the connection is only defined up to local 1-forms $\chi$ on $\A$. 
This makes the Schwinger term defined up to terms $\delta \chi $. The 
consistent Schwinger term is thus the curvature corresponding to the canonical 
connection, up to local forms, on the vacuum bundle. The relation $\delta 
F=0$, due to the Bianchi identity, gives the consistency condition for the 
Schwinger term. This can be seen directly from the fact that
\begin{eqnarray*}
\delta \int _M\omega _{2n+1 }(A+v,A\pow {\prime }) & = & \int _M(d+\delta )\omega 
_{2n+1 }(A+v,A\pow {\prime }) \nonu 
& = & \int _M\left(
\mbox{tr}\left( dA+A\pow 2\right) \pow {n+1} -\mbox{tr}\left( dA\pow {\prime 
}+A\pow {\prime 2}\right) \pow {n+1}\right) =0
\end{eqnarray*}
holds in gauge directions. As for the consistent anomaly, it implies that 
the consistent Schwinger term can be regarded as an element in a cohomology 
group.

If $t$ is interpreted as time, then the consistent Schwinger term has been 
computed by comparing the vacuum bundle at the time of interest ($t=1$) with 
a reference vacuum bundle at another time ($t=0$). The reference vacuum is 
given by the Hamiltonian $H_{A\pow {\prime }}$. Certainly, it is possible to 
use the field $A$ itself as a background, i.e. as for the anomaly we now put 
$A\pow {\prime }=A$. In this case eq. \Ref{eq:6} gives the covariant 
Schwinger term. Thus, the figure in the previous section illustrates as well 
the case of Schwinger terms (except for the fact that the line bundle over (a subset of) $\A\times\A$ is  different).

That both the consistent and covariant anomaly are given by connections 
suggests that both the consistent and covariant Schwinger term are given by 
curvatures. For the covariant Schwinger term this is however not true in 
general. This is easy to understand since all curvatures of line bundles are 
closed by the Bianchi identity. Thus, if the covariant Schwinger term could 
have been obtained from a curvature, then it would also have to satisfy the 
consistency condition. This is however not true. The reason why the covariant Schwinger term in 
general is not obtained from a curvature is that the third term on the right 
hand side of eq. \Ref{eq:5} is not necessary zero in gauge directions for 
$A\pow {\prime }=A$. This term is determined by the boundary conditions at 
the manifold $M\times\{ t=0 \} $. Since we are interested in the other part 
$M\times\{ t=1 \} $ of the boundary when computing the Schwinger term, it is 
reasonable to not take into account the contribution from $\hat{\eta 
}_{\lambda \pow {\prime }}$. We thus only pay attention to how the connection of the universal bundle at $M\times \{ t=1\}$ is extended to the bulk and not to terms coming 
from data on the other part of the boundary. Let us give an alternative 
description, intended for readers unfamiliar with the $\hat{\eta }$-forms. 
Extend $M$ to a manifold which have $M$ as its boundary and looks like a 
cylinder close to $M$, i.e. the extension is $M\times I\cup\tilde{M}$, where 
$\partial \tilde{M}=M\times \{ t=0\} $. Then extend the gauge connections 
and gauge transformations in a smooth way to the bulk. The curvature of the 
corresponding bundle is $F+\tilde{F}$, where $F$ is given by eq. \Ref{eq:5} 
and $\tilde{F}$ is defined with respect to $\tilde{M}$. If we consider the 
gauge connection $A(t=0)$ as a fixed background field $A \pow {\prime }$ 
which doesn't depend on $A=A(t=1)$, then $\tilde {F}$ is zero and the consistent 
anomaly is obtained. If $A(t=0)=A(t=1)$, then
\[
\tilde{F}=c_n\left(\int _{\tilde{M}}\mbox{tr}
\left( \tilde{\F }\pow 
{n+1}\right) -\frac{1}{2}(\hat{\eta }_{\lambda \pow {\prime }})_{[2]}\right)
\]
where $\tilde{\F }=(d+\delta )A+A\pow 2$. It gives 
\[
F+\tilde{F}=c_n\left(\int_{M\times I}\mbox{tr}\left( \F \pow 
{n+1}\right) +\int _{\tilde{M}}\mbox{tr}
\left( \tilde{\F }\pow 
{n+1}\right) -\frac{1}{2}(\hat{\eta }_{\lambda })_{[2]}\right) .
\]
As in the consistent case, $\hat{\eta }_{\lambda }$ gives zero in gauge directions. The contribution from the second term comes from a manifold which is far from the manifold $M\times \{ t=1\}$ of interest. For this reason, it can be disregarded. The same is true for the second term on the left hand side.

\begin{center}\epsfxsize=4cm
$$\hbox{\epsfbox{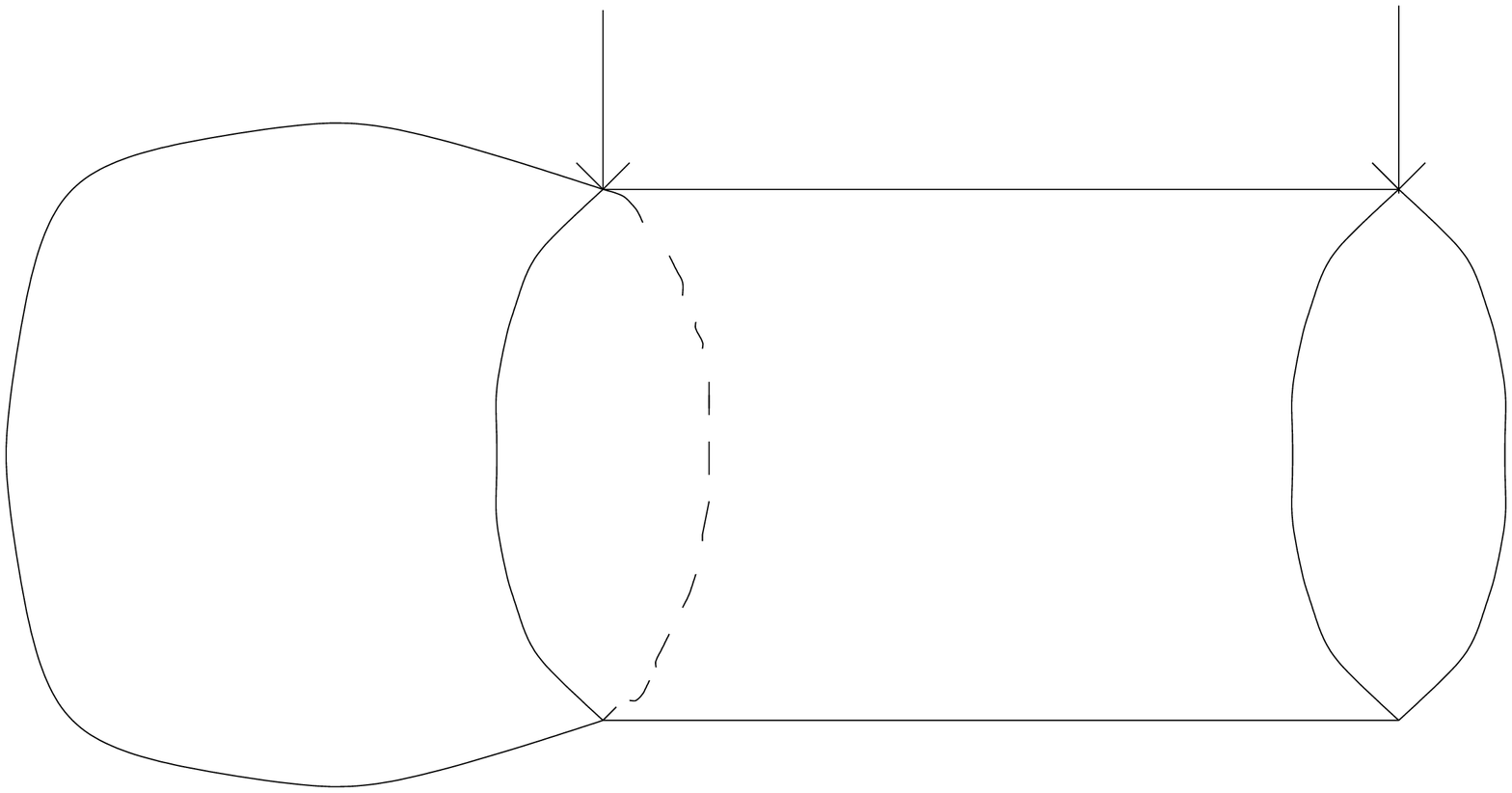}}$$
\end{center}
\begin{picture}(0,0)(0,0)
\put(195,90){\footnotesize{Schwinger term}}
\put(135,90){\footnotesize{Background}}
\put(155,15){\footnotesize{$t=0$}}
\put(210,15){\footnotesize{$t=1$}}
\put(180,15){\footnotesize{$M\times I$}}
\put(130,10){\footnotesize{$\tilde{M}$}}
\end{picture}

The literature about covariant Schwinger terms can be divided into two parts depending on if they define it by an algebraic argument \cite{TS,K1} or by a physical argument, see for instance \cite{HS,NT,K2}. The first discipline have a lack of physical understanding while the mathematical structure is unclear in the second. In this paper, on the other hand, a mathematical rigorous construction of an underlying geometrical and physical idea has been made. It is therefore not sound to use earlier results as a judge if the non-local $\hat{\eta }$-form will give a \pagebreak contribution or not. Indeed, all the previous computations of the covariant Schwinger term seems already from the beginning to neglect the possibility of contributions from terms as the non-local $\hat{\eta }$-form. From this point of view it seems more natural to define the covariant Schwinger term as the curvature above. We then obtain a non-local covariant Schwinger term which differs from earlier results by $\frac{1}{2}(\hat{\eta }_{\lambda \pow {\prime }})_{[2]}$. However, since everything comes down to the definition we will for the rest of the paper disregard the  $\hat{\eta }$-form so that our result agrees with the previous ones. 

Notice that the discussion above also makes sense for the anomaly in the space-time formalism: With the mathematical methods of section 2 we obtain a contribution from an $\hat{\eta }$-form when the space-time has a boundary.

\section{Descent equations and connecting terms}
Let $\omega \pow k_{2n+1-k}(A+v,A\pow \prime )$ denote the part of $\omega _{2n+1}(A+v,A\pow \prime )$ which have ghost degree $k$. Expansion in ghost degrees of the first expression in \Ref{eq:2} with $\alpha =A+v$ and $\alpha \pow \prime =A\pow \prime $ gives the consistent descent equations:
\begin{eqnarray*}
\mbox{tr} \left( dA +A\pow 2\right) \pow {n+1} -\mbox{tr} \left( dA\pow \prime +A\pow {\prime 2}\right) \pow {n+1} & = & d\omega \pow 0_{2n+1}(A+v,A\pow \prime )\nonu
\delta\omega \pow k_{2n+1-k}(A+v,A\pow \prime ) & = & -d\omega \pow {k+1}_{2n-k}(A+v,A\pow \prime ), \nonu && \quad k=0,1,...,2n\nonu
\omega \pow {2n+1}_{0}(A+v,A\pow \prime )=0,
\end{eqnarray*}
where it has been used that $\F =dA+A\pow 2$ in gauge directions. From the previous sections it follows that the (non-integrated) anomaly and Schwinger term in a background $A\pow \prime$ is $-c_n \omega \pow 1_{2n}(A+v,A\pow \prime )$ respective $-c_n \omega \pow 2_{2n-1}(A+v,A\pow \prime )$. We thus see that the descent equations can be used for their computation. Further, it gives a relation between them. By integrating over $M$ it is also seen that the consistency condition is fulfilled. When $A\pow \prime =0$ we obtain the \lq ordinary\rq $ $ descent equations. 

It is not so well-known that there exist descent equations for the covariant anomaly and Schwinger term as well. The first relation in \Ref{eq:2} with $\alpha =A+v$ and $\alpha \pow \prime =A$ gives the covariant descent equations \cite{K1}: 
\begin{eqnarray*}
0 & = & \omega \pow 0_{2n+1}(A+v,A)\nonu
\delta\omega \pow k_{2n+1-k}(A+v,A) & = & -d\omega \pow {k+1}_{2n-k}(A+v,A) \nonu &&- \left( \begin{array}{c} n+1\\ k+1 \end{array}\right) \mbox{str}\left( (\delta A)\pow {k+1} (dA+A\pow 2)\pow {n-k}\right),\nonu
&& \quad k=0,1,...,n\nonu
\delta\omega \pow k_{2n+1-k}(A+v,A) & = & -d\omega \pow {k+1}_{2n-k}(A+v,A) ,\quad k=n+1,...,2n\nonu
\delta\omega \pow {2n+1}_{0}(A+v,A) & = & 0,
\end{eqnarray*}
where $\mbox{str}$ is the symmetrized trace. This gives a computational scheme for the covariant anomaly and Schwinger term. However, \pagebreak as for the consistent case,
  the direct formula, eq. \Ref{eq:2}, is to prefer. Let us give the explicit expressions in the descent equations for the simplest case $n=1$: 
\begin{eqnarray*}
\delta 2\mbox{tr}\left( v(dA+A\pow 2)\right) & = & -d\mbox{tr}(v\delta A)- \mbox{tr}(\delta A)\pow {2} \nonu
\delta \mbox{tr}(v\delta A) & = & -d\left( -\frac{1}{3}\mbox{tr}v\pow 3\right)\nonu
\delta\left( -\frac{1}{3}\mbox{tr}v\pow 3\right) & = & 0,
\end{eqnarray*}

The triangle formula \cite{MSZ} implies that $\omega \pow k_{2n+1-k}(A+v,A)$ is up to trivial terms (which in the non-integrated setting takes the form $(d+\delta )\chi$) equal to $\omega \pow k_{2n+1-k}(A+v,A\pow \prime )-\omega \pow k_{2n+1-k}(A,A\pow \prime )$. The difference between the consistent and covariant cochains is thus given by $\omega \pow k_{2n+1-k}(A,A\pow \prime )$. When $k\geq n+1$ this term is zero and the consistent and covariant cochains are equal. Certainly, this is in agreement with the fact that the consistent and covariant descent equations are equal in this case. Thus, half of the cochains are always equal. It explains the fact that the consistent Schwinger term is covariant when $n=1$, while this is not true for $n\geq 2$. 

The connecting term $ \omega \pow k_{2n+1-k}(A,A\pow \prime )$ has a very interesting property. It contains only the ghost in the combination $\delta A$. This implies that it can be extended to a local form on all of $\A$, not necessary in gauge directions. (With local we mean that it can be expressed as a trace of a polynomial in $A$, $dA$ and $\delta A$.) This is in contrary to $ \omega \pow k_{2n+1-k}(A+v,A\pow \prime )$ and $ \omega \pow k_{2n+1-k}(A+v,A)$ since if the ghost is extended to all of $\A$, for instance to $(d_A^\ast d_A)^{-1}d_A^\ast $, then a non-local expression is obtained. 

That the covariant cochains are not in any cohomology group has sometimes been regarded as a \lq drawback\rq $ $ of the covariant formalism. This is not really true. In fact, the covariant cochains can be regarded as elements in a cohomology group due to their bijective correspondence with such elements, namely the consistent cochains. In general, since they are in a one-to-one relation, the consistent and covariant terms are always on equal footing. Another example where this can be used is in the fact that the consistent anomaly and Schwinger term can be pushed forward to $\A /\G$. These geometrical objects on line bundles over (subsets of) $\A$ becomes topological on $\A /\G$. That this is not possible for the corresponding covariant terms follows from the fact that $\widehat{\mbox{DET}}i(\partial \!\!\! /_A \pow ++\partial \!\!\! /_{A\pow \prime } \pow -)$ can be pushed forward to $\A /\G$ while this is not true for $\widehat{\mbox{DET}}i(\partial \!\!\! /_A \pow ++\partial \!\!\! /_{A} \pow -)$.

We would also like to point out that the chiral anomaly and the Schwinger term (and the other cochains) can appear in different forms than just the consistent and covariant. In general, the connection $A\pow \prime $ can be separated into a piece $f(A)$ which is independent of $A$ and a piece $h(A)$ which depends on $A$. The extreme case $h(A)= A\pow \prime $, $f(A)=0$ gives the consistent cochains while $h(A)= 0$, $f(A)=A$ gives the covariant cochains.

\thanks{\bf Acknowledgments:} I thank Dr C. Adam for an interesting discussion.


\newpage
\begin{thebibliography}{}
\bibitem{AES} Adam, C., Ekstrand, C., Sykora, T., in preparation
\bibitem{BGV} Berline, N., Getzler, E., Vergne, M.,  
{\sl Heat Kernels and Dirac Operators}, Springer-Verlag,
Berlin Heidelberg New York, 1992
\bibitem{BF} Bismut, J. -M., Freed, D. S., Commun. Math. Phys. {\bf 106}, 159 (1986); {\bf 107}, 103 (1986)
\bibitem{AS} Atiyah, M. F., Singer, I. M.,
 Proc. Nat. Acad. Sci. 
{\bf 81}, 2596 (1984)
\bibitem{MSZ} Ma\~{n}es, J., Stora, R., Zumino, B., Commun. Math. Phys. {\bf 102}, 157 (1985)
\bibitem{KO} Kontsevich, M., Vishik, S., hep-th/9406140
\bibitem{CMM} Carey, A. L., Mickelsson, J., Murray, M.K., 
     Comm. Math. Phys. {\bf 183}, 707 (1997)
\bibitem{APS} Atiyah, M. F., Patodi, V. K., Singer, I. M., Math. Proc. Camb. Phil. Soc.  
{\bf 77}, 43 (1975)
\bibitem{PZ2} Piazza, P., J.  Commun. Math. 
Phys. {\bf 193}, 105 (1998)
\bibitem{EM} Ekstrand, C., Mickelsson, J., to appear in Comm. Math. Phys.
\bibitem{TS} Tsutsui, I., Phys. Lett. B {\bf 229}, 51 (1989)
\bibitem{K1} Kelnhofer, G., J. Math. Phys. {\bf 34}, 3901 (1993)
\bibitem{HS} Hosono, S. and Seo, K, Phys. Rev. D {\bf 38}, 1296 (1988)
\bibitem{NT} Nishikawa, T., Tsutsui, I., Nucl. Phys. B {\bf 308}, 544 (1988)
\bibitem{K2} Kelnhofer, G., Z. Phys. C {\bf 52}, 89 (1991)
\end{thebibliography}
\end{document}